\documentclass[aps, prl, twocolumn, showpacs, superscriptaddress, floatfix]{revtex4-1}
\usepackage{graphics,amssymb,amsmath,epsfig,color}
\newcommand{\be}{\begin{equation}}
\newcommand{\ee}{\end{equation}}
\newcommand{\bea}{\begin{eqnarray}}
\newcommand{\eea}{\end{eqnarray}}

\begin{document}

\title {Comment on ``Dynamic Opinion Model and Invasion Percolation"}
\author{Arsalan Sattari, Maya Paczuski, and Peter Grassberger} \affiliation{Complexity Science Group, University of Calgary, Calgary, Canada}


\date{\today}

\maketitle

In~\cite{shao2009} Shao {\it et al.} claim, based on low statistics simulations, that 
a model with majority rule coarsening exhibits in $d=2$ a 
percolation transition in the universality class of invasion 
percolation {\em with trapping} (IPT).
They report also that the system reaches its final state rapidly 
with no diverging time scale. Since the original configurations are random and thus 
in the ordinary percolation (OP) universality class, it seems unlikely 
that long range correlations could develop in a finite time that would 
change this. Indeed it was proved rigorously~\cite{camia2004} that 
similar 2-d models (called ``dependent percolation" in~\cite{camia2004}) belong to the 
OP universality class.  

Here we present high statistics (up to $L=2^{14}$, $>10^4$ realizations)
on $L\times L$ square lattices and confirm that the phase transition 
is in the OP universality class, thus refuting a central tenet of~\cite{shao2009}.
Initially each site $i$ is randomly assigned one of two opinions (or spins): 
$\sigma_i=+1$ with probability $f$, otherwise $\sigma_{i}=-1$. At each time step all 
sites are updated in parallel. If at least three of their four neighbors disagree 
with them, they change their opinion, otherwise they keep it. As noted
in~\cite{shao2009} this leads quickly (within $O(10)$ time steps) to a static state, 
except for sites that flip permanently with period 2. The critical probability $f_c$ 
where a ``+1" cluster percolates depends slightly on how these flicker sites are 
treated (we treat them as ``+", if $\sigma_i=+1$ at even times), but the universal 
properties do not.

We first determine $f_c$ by measuring the chance that a cluster in the final state percolates 
through lattices with open boundary conditions. Using finite size scaling~\cite{ziff2002}, we 
obtain $f_c = 0.506425(20)$, in agreement with the less precise estimate of~\cite{shao2009}. 
After that, 
we measure the distribution of cluster sizes with $\sigma=+1$ in final states obtained with
helical b.c. for $f\approx f_c$. Figure 1 confirms the above estimate of $f_c$ and shows that 
the data are excellently described by a power law $P(s) \sim s^{-\tau}$\ with the OP critical exponent 
$\tau=187/91\approx 2.055$, ruling out the IPT exponent $\tau\approx 1.89$. We see also deviations
from this power law at small masses $s$, as small clusters are eliminated by the coarsening.  This, 
together with using open boundary conditions and neglecting finite size corrections, explains 
why $\tau$ was underestimated in~\cite{shao2009}.
A data collapse of the r.h.s peaks in Fig.1 gives $D_f = 1.895(15)$ as for OP, but in disagreement
with IPT. Notice that the exponents obtained in~\cite{shao2009} strongly violate the hyperscaling 
relation $\tau = d/D_f+1$.

Shao {\it et al.} claim that IPT is relevant because local clusters get trapped. The 
difference between OP and IPT is that clusters can grow both outwards and inwards 
(into empty holes) in OP, while they can only grow outward in IPT. In this respect the 
model of~\cite{shao2009} is exactly as OP.

We also simulated the process on random Erd\"os-Renyi networks. For small average 
degrees we confirm the claim of~\cite{shao2009} that the percolation transition is in the 
OP class. But for large average degrees we find an unexpected first order 
transition \cite{arsalan2012}.

As a model for opinion dynamics the model is of limited interest, since the dynamics leaves essentially
unchanged all large clusters present in the initial state -- except for clusters with hubs, in case of 
scale-free networks, who immediately adopt the majority opinion. 

\begin{figure}[h]
\begin{center}
\vskip -4pt
\psfig{file=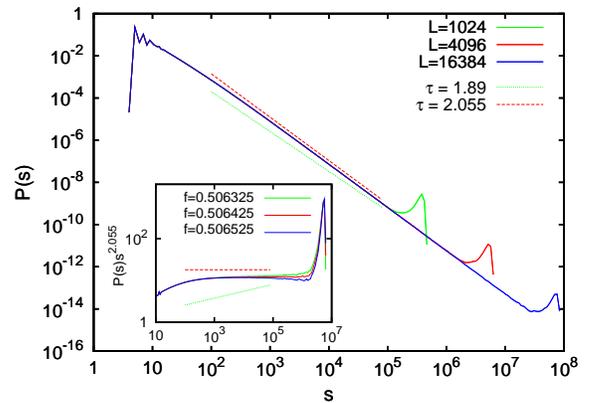,width=5.5cm, angle=270}
\vskip -7pt
\caption{(Color online) Probability distribution of cluster sizes at $f=0.506425$ for different $L$ on a 
   log-log scale. The straight lines represent power laws with exponent $\tau=2.055$ (red, dashed) 
   and $\tau=1.89$ (green, dotted), corresponding to OP and IPT, respectively. Inset: Data for $L=4096$ 
   at different values of $f$, after multiplication with $s^{2.055}$. Small 
   changes of $f$ give rise to deviations from power law behavior. At our estimate of $f_c$ obtained
   independently from the spanning probability, power law scaling with the OP exponent $\tau=2.055$ 
   extending over three orders of magnitude.}
\vskip -16pt

\label{fig1}
\end{center}
\end{figure}

We thank a referee for pointing out the violation of hyperscaling in~\cite{shao2009}.
\bibliographystyle{apsrev4-1}
\bibliography{shaobib}
\end{document}